%% file: main.tex
\documentclass[%
 reprint,
 superscriptaddress,
 amsmath,amssymb,
 aps, prd,
floatfix,
]{revtex4-2}

\usepackage{graphicx}
\usepackage{dcolumn}
\usepackage{bm}

\usepackage[mathlines]{lineno}

\usepackage{upgreek} 

\usepackage{xspace}
\usepackage[dvipsnames]{xcolor}
\usepackage{multirow}

\newcommand{\Ricochet}{\textsc{Ricochet}\xspace}
\newcommand{\cenns}{CE$\nu$NS\xspace}

\begin{document}
\title{Characterization of mini-CryoCube detectors from the RICOCHET experiment commissioning at the Institut Laue-Langevin}
\input{authors}
\date{\today}

\begin{abstract}
The \Ricochet experiment aims to measure the coherent elastic neutrino-nucleus scattering process from  antineutrinos emitted by a research nuclear reactor operated by the Institut Laue-Langevin (Grenoble, France). 
This article presents a description of the \Ricochet experimental installation and the detector performance achieved during its commissioning with a mini-CryoCube module consisting of three 42-gram germanium cryogenic calorimeters. 
The baseline resolutions and background levels are reported both during reactor-on and reactor-off periods, and as noise mitigation techniques were improved.
A baseline resolution of 40~eV electron equivalent was achieved for the ionization channel after setup improvements, and the phonon channel resolutions ranged from 50 to 80~eV of total phonon energy. 
In the energy region from 2 to 7~keV, a nuclear recoil rate of 15(2)~events/(kg\,day\,keV) is measured during the reactor-off period selecting events in coincidence with muon veto signals. 
This rate is in agreement with the cosmogenic neutron rate calculated from GEANT4 simulations. 
After the rejection of events in coincidence with signals in the muon veto detectors, a combined 90\%~C.L. limit on the nuclear recoil background of $<9$~events/(kg\,day\,keV) is obtained in that energy region during the reactor-on period, which is compatible with our GEANT4 model calculation corresponding to a total rate of 5~events/(kg\,day\,keV). 
The sensitivity of this analysis was however found to be limited by a surface event contamination which is currently being addressed by the \Ricochet Collaboration with upgraded detectors.
\end{abstract}

\maketitle

\section{Introduction}
\label{sec:intro}

The coherent elastic neutrino-nucleus scattering (\cenns) process was proposed by Freedman in 1974 as a new gateway to study neutrino interactions~\cite{Freedman:1974}.
Forty-three years later, the COHERENT Collaboration~\cite{Akimov:2017,Akimov:2021,Akimov:2022a,Adamski:2024} measured it for the first time at the spallation neutron source (SNS) at Oak Ridge National Laboratory.
These first detections with CsI[Na], argon and germanium detectors opened a new measurement channel to probe the neutrino sector for physics beyond the Standard Model.
The \cenns spectral shape at low momentum transfers may be affected by various beyond Standard Model physics scenarios~\cite{Lindner:2017, Billard:2018, Billard:2021, Cadeddu:2018, Corona:2025} including: non-standard interactions of neutrinos and quarks, neutrinos coupling to new massive scalar and vector mediators, anomalies of the neutrino magnetic moment, neutrino charge radius, or oscillations between active and sterile neutrinos.
Since a high-statistics and high-precision measurement of the \cenns process is required to explore physics beyond the Standard Model, a high neutrino flux is desirable. 
Nuclear reactors can provide a more intense neutrino flux than the SNS (four orders of magnitude higher for a research reactor~\cite{Akimov:2022,Augier:2023}), with the challenge of a factor of 10 lower neutrino energy. 
However, the lower energy has the important benefit that the \cenns process is fully coherent, so that nuclear form factors approach unity~\cite{Sierra:2019}. 

Possible hints of observations of \cenns with high-purity germanium detectors at nuclear reactors were reported by Dresden-II~\cite{Colaresi:2021} and CONUS+~\cite{Ackermann:2025}. 
For experiments aiming to detect \cenns of reactor antineutrinos, the current challenge is achieving a low energy threshold which both strongly increases the \cenns event number~\cite{Lindner:2017} and thus the statistical significance of a measurement, and is needed to explore new physics scenarios at low energy. 
With their capacity for low threshold and particle identification, cryogenic calorimeters are well suited for exploring the \cenns process at reactors.

The \Ricochet experiment~\cite{Augier:2023a} aims to measure \cenns with an array of eighteen 42-gram germanium cryogenic calorimeters at the Institut Laue-Langevin (ILL) in Grenoble (France).
This array (called CryoCube) will be arranged in two floors of nine detectors each, which will be divided into independent three-detector modules called mini-CryoCubes~\cite{Augier:2024}.
The cryogenic experimental setup of \Ricochet was installed at the end of 2023 at the ILL. 
The cryogenic performance was validated in a week-long run~\cite{run12:2014} without detectors in early February 2024, where a base temperature of 8.6~mK was achieved.
This article presents the results of the following two commissioning runs spanning two reactor-on and reactor-off cycles and employing a mini-CryoCube module: (1)~RUN013~\cite{run13:2014} from February~19$^{\mathrm{th}}$ to April~4$^{\mathrm{th}}$, 2024; (2)~RUN014~\cite{run14:2014} from May~7$^{\mathrm{th}}$ to October~14$^{\mathrm{th}}$, 2024.
The latter measurement provides detailed feedback on the types of backgrounds that may limit the sensitivity of the experiment in the future and leads us to the identification of mitigation. 
The article is structured as follows: the experimental setup and data collection strategy are described in Sec.~\ref{sec:setup}. 
The data processing and analysis are reported in Secs.~\ref{sec:processing} and~\ref{sec:analysis}, respectively.
The detector resolution performance is presented in Sec.~\ref{sec:performance} and the background levels achieved are discussed in Sec.~\ref{sec:bkg}. The conclusions and outlook are summarized in Sec.~\ref{sec:conclusions}.

\section{Experimental setup and data collection} 
\label{sec:setup}
This section describes the cryogenic facility in Sec.~\ref{sec:facility} and the cryogenic calorimeters in Sec.~\ref{sec:det}. A summary of the data collected during commissioning is reported in Sec.~\ref{sec:data}.
\subsection{Cryogenic facility}
\label{sec:facility}
The \Ricochet experiment is hosted at the research nuclear reactor of the ILL, which has a nominal thermal power of 58~MW. 
The experiment is located at the H7 site, with the \Ricochet detectors positioned 8.8~m from the center of the reactor core. 
The site provides shielding from cosmic rays corresponding to an overburden of $\sim$15~m water equivalent~\cite{Allemandou:2018}, given both by the building and the transfer channel of the reactor, a concrete channel filled with water, which covers most of the experiment.
Typically, the ILL provides three reactor cycles per year of $2.6\times10^3$ MW\,day each, where the duration of one cycle is between 45 and 63~days. 
The comparison of data in reactor-on and reactor-off cycles allows us to subtract non-reactor correlated background, and the multiple cycles allow us to closely monitor the variation of the background as a function of time.

The \Ricochet detectors are operated in a CryoConcept~\cite{Cryoconcept:2021} Hexa-Dry 200 dilution refrigerator equipped with ultra quiet technology (UQT) to minimize noise induced by vibrations from the pulse tube.
Cryoconcept's UQT is implemented by mounting the dilution refrigerator and the pulse-tube system (cold head, rotating valve, and gas ballasts) on two different triangular frames (called the cryostat frame and the pulse-tube frame, respectively).
The thermal exchange between the cold head and the fridge plates is achieved by convection using the $^3$He/$^4$He mixture within the dilution circuit.
The significant reduction of mechanical contact between the pulse tube and the cryostat, which is limited to only a gas-sealing bellows with low natural vibration frequency, minimizes the propagation of vibrations~\cite{Olivieri:2017}.

\begin{figure*}[t]
    \includegraphics[width=0.3\textwidth]{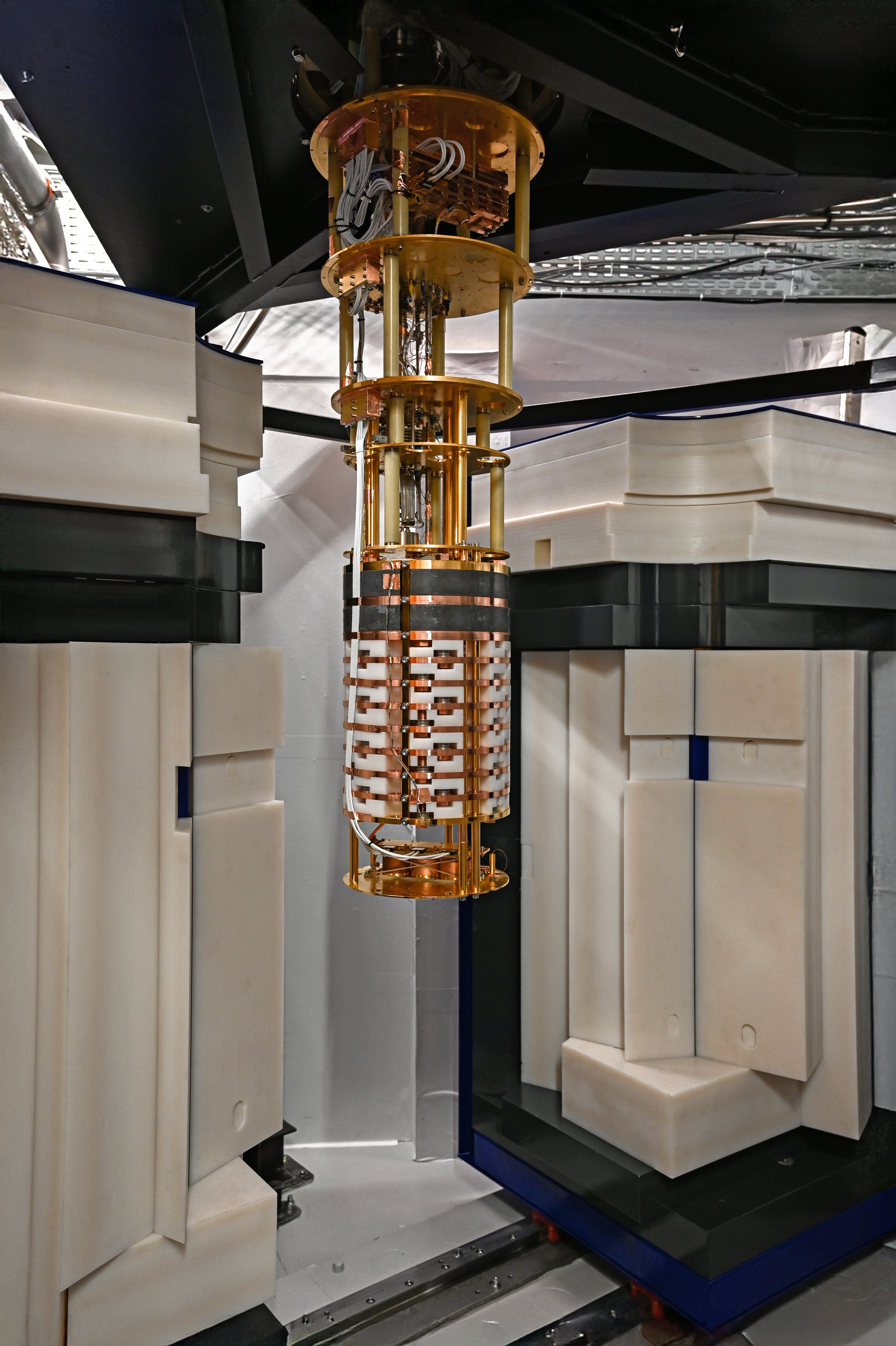}
    \includegraphics[width=0.68\textwidth]{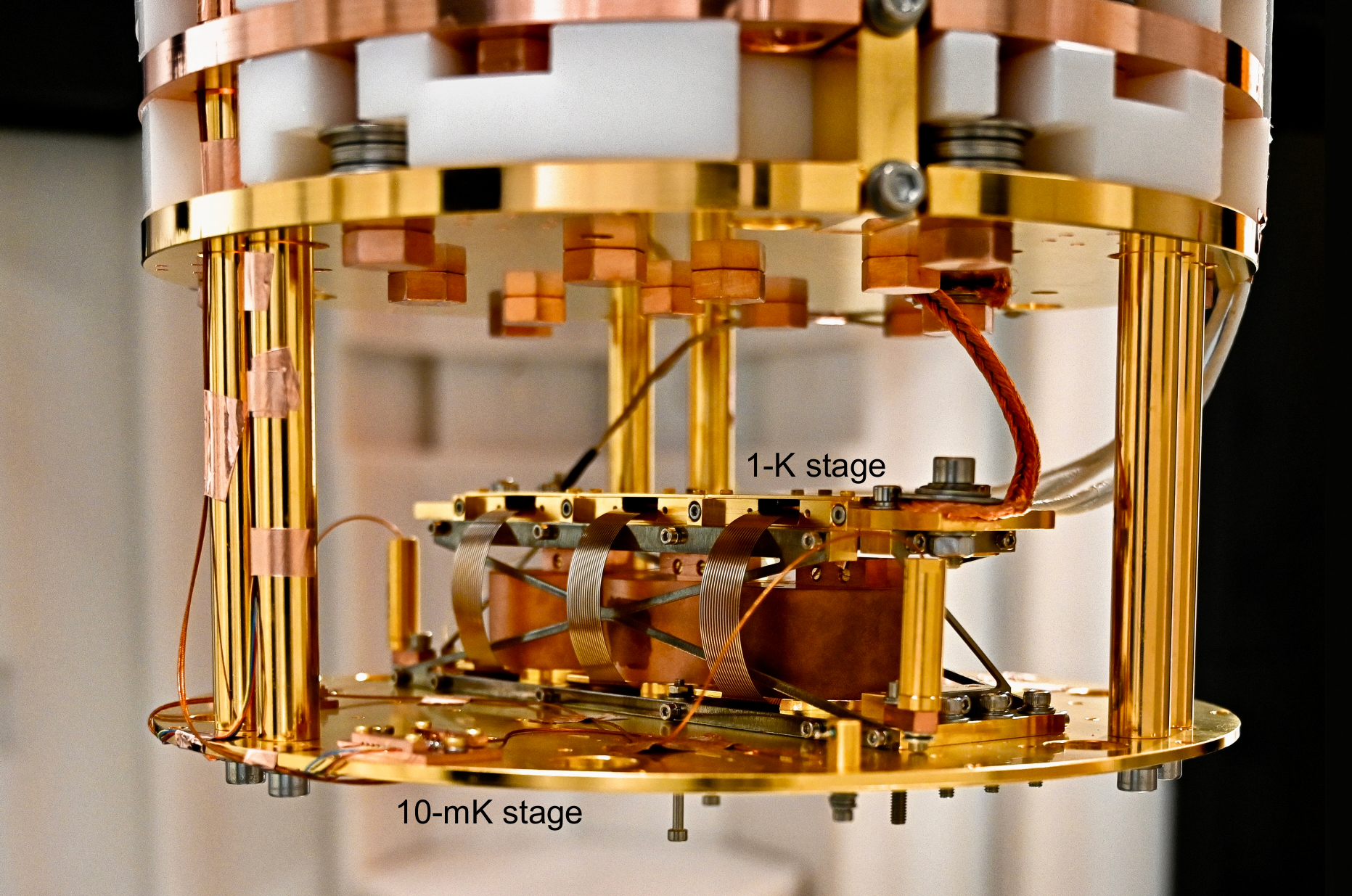}
    \caption{Left panel: photograph of the \Ricochet cryostat installed at the ILL on its triangular frame during RUN014. The inner shielding is mounted at the 1-K stage with HDPE layers interposed between copper disks for thermalization. The two dark gray disks are made of lead. 
    A mini-CryoCube was placed below the inner shield at the 10-mK stage. Two segments of the outer shield are visible in the background: the white parts are the borated HDPE and the black parts are the lead. Right panel: photograph of the mini-CryoCube mounted on the \Ricochet cryostat. The germanium detectors are covered by a cylindrical copper hat. They are mounted on the 10-mK stage. The HEMT stage is placed above the copper hats and is thermalized at 1~K through a copper braid. The HEMT amplifier stage is supported by two x-shaped and two z-shaped structures made of a titanium alloy and are connected through Kapton-based flex PCB with CuNi leads to the detectors.}
    \label{fig:photo}
\end{figure*}

The cryostat is surrounded by a 22-ton external shield to limit reactogenic and radiogenic backgrounds. It is composed of two parts: (1)~a cup-shaped shield around the detector area which consists of a 35-cm-thick inner layer of borated high-density polyethylene (HDPE) and a 20-cm-thick outer layer of lead; (2)~two top layers of 20-cm-high lead (bottom) and 35-cm-high HDPE (top) that further attenuate neutrons and backgrounds coming from above.
The external shield is enclosed in a soft iron shell, which attenuates the magnetic fields generated by the neighboring experiments, and is mounted on a rail system to ease access to the cryostat. More details about the external shield can be found in Ref.~\cite{Augier:2023a}.
Figure~\ref{fig:photo} (left) shows a photograph of the cryostat installed at the ILL, where the outer and inner shielding are visible.

The external shield does not cover the top of the cryostat. For this reason, an internal shielding (Fig.~\ref{fig:photo}, left) was designed to fill the cylindrical space above the detectors. 
The internal shield consists of two parts: (1)~a cylindrical plug mounted at the 1-K stage, composed of two 4.25-cm-high lead disks inserted between three 1.5-cm-high copper disks, followed by 16~layers of alternating $\sim$3.5-cm-high HDPE disks and 1-cm-high copper disks; 
(2)~8-mm-thick HDPE sheets installed around the copper cans of the 1-K, 4-K and 50-K stages at the height of the internal plug.
Both the cylindrical plug and the HDPE sheets were installed between RUN013 and RUN014.

During RUN014, an active outer muon veto was installed around the external shield to tag muons and related cosmogenic backgrounds. 
The outer muon veto consists of 34~plastic scintillator panels (200$\times$50$\times$3~cm$^3$) repurposed from the GERDA experiment~\cite{Ackermann:2013}.
The panels are arranged in 17 pairs to form an inner and outer layer. The coincidence signals of two panels allow us to reject triggers from reactogenic $\gamma$ backgrounds.
Six pairs, resized from GERDA in length and shape, are placed on top of the outer shielding and eleven pairs encircle its sides.
Each muon veto panel is read out via a single photomultiplier tube (Hamamatsu R960) located at one extremity of the panel, the light collection is ensured by optical fibers glued on the panel's lateral faces. The signals are read out by a custom electronic system~\cite{Bourrion:2016}. The muon veto panels use a hardware trigger that records signals from all panels when a certain threshold is surpassed, in at least one inner and one outer muon veto panel.
The resulting muon rate was roughly 500~Hz in RUN014.
A dedicated electronics board provides the same clock to the data acquisition systems of the muon veto and the cryogenic calorimeters, ensuring synchronization of the two data streams.

The \Ricochet site is located above a pump system which operates when the reactor is on, inducing vibrations in the floor.
During the commissioning, a series of upgrades were made to decouple the cryostat from floor vibrations.
Damping pads~\cite{Fabreeka:2019} were installed between the cryostat frame and its pillars in RUN013 to attenuate the vibrations propagating from the floor. 
The decoupling of the pulse tube was also improved. First, the pulse-tube gas ballasts and rotating valve were moved to a third support structure fixed to the ceiling during RUN013. Second, a sorbothane-rubber strip was added around the hose connecting the pulse-tube rotating valve to the cold head during RUN014.
Since the pulse-tube frame is rigidly bolted to the floor, the floor vibrations may still propagate to the cryostat frame via the soft bellows connecting the two frames.
Therefore, the section of the pulse-tube frame connecting to the bellows was interposed between two damping polyurethane foams~\cite{Regufoam:2018} in RUN014. 
The impact of the installation of the different damping elements on the baseline energy resolutions is discussed in Sec.~\ref{sec:performance}.

\subsection{Cryogenic calorimeters}
\label{sec:det}
The mini-CryoCube module~\cite{Augier:2024} tested during the commissioning is shown in Fig.~\ref{fig:photo} (right). It was composed of three 42-gram germanium detectors (RED167, RED237 and RED137). 
These detectors are instrumented with both phonon and ionization readout~\cite{Armengaud:2017, Salagnac:2023} to distinguish  nuclear recoil signals from backgrounds due to $\gamma$ and $\beta$ radiation, as well as heat-only events~\cite{Armengaud:2022, Adari:2022, Baxter:2025}.
The phonon channel is read out with a germanium neutron-transmutation-doped (Ge-NTD) thermistor~\footnote{The Ge-NTD sensors from the EDELWEISS-III experiment~\cite{Armengaud:2017} were reused and adapted to meet the specifications of the CryoCube design.} and the ionization channel is read out with two aluminum electrodes in a planar configuration~\cite{Augier:2024}.
A bias current of 0.5~nA was chosen for the Ge-NTD thermistors in order to optimize the detector baseline resolution when the detectors were thermalized at 14.2~mK. 
A symmetric voltage of $\pm 2$~V was applied across the electrodes for the ionization readout, which amplifies the phonon channel signal via the Neganov-Trofimov-Luke (NTL) effect~\cite{Neganov:1985, Luke:1988}. 
One of the two ionization channels of RED137 was not operational, therefore the data taking focused on the other two detectors (RED167 and RED237).  

The energy measured by the detector will be expressed by three different quantities: 
\begin{itemize}
\item the total phonon energy ($E_{ph}$) is expressed in units of keV$_\mathrm{ph}$ when the signal is composed of the initial recoil energy ($E_{r}$) summed with the NTL phonon amplification ($E_{NTL}$): 
\begin{equation}
\label{eqn:boost}
    E_{ph} = E_r + E_{NTL} = E_r + E_r \cdot \mathcal{Q} \dfrac{q \cdot V}{\epsilon} ,
\end{equation}
where $q$ is the electron charge, $V$ is the voltage bias applied across the detector target and $\epsilon$ is the average energy required to produce one electron-hole pair, assumed equal to 3~eV for electron recoils in cryogenic germanium detectors~\cite{Knoll:2010, Chapellier:2000}.
The quenching between electron and nuclear recoils is taken into account by the ionization yield $\mathcal{Q}$, which is equal to 1 for electron recoils and depends on the recoil energy for nuclear recoils with typical values around 0.2 for keV-scale recoils~\cite{Jones:1971, Jones:1975, Messous:1995, Benoit:2001, Benoit:2007, Barbeau:2007, Armengaud:2017, Bonhomme:2022};
\item the ionization channel energy ($E_{ion}$) has keV$_\mathrm{ee}$ as units, where keV$_\mathrm{ee}$ (keV electron equivalent) corresponds to the measured charge equivalent to that generated by an electron recoil event with an energy of 1~keV;
\item the reconstructed recoil energy is deduced from $E_{ph}$ and $E_{ion}$ using Eq.~\ref{eqn:boost} and the fact that $E_{ion} = \mathcal{Q} \cdot E_r$:
\begin{equation}
\label{eqn:recoil}
    E_r = E_{ph} - E_{ion} \dfrac{q \cdot V}{\epsilon}.
\end{equation}
The reconstructed recoil energy is in keV units.
\end{itemize}

The first stage of amplification for the phonon channels is performed with field-effect transistors (FETs) installed on the 50-K stage of the cryostat through a thermal decoupler to work in a temperature range between 80 to 120~K. For the ionization channels, high electron mobility transistors (HEMTs)~\cite{Juillard:2020,Dong:2014} operated at 1~K are used for amplification. The HEMTs are mounted on top of the germanium detectors (Fig.~\ref{fig:photo}, right) with titanium-alloy x-shaped and z-shaped structures~\cite{Augier:2024}.
The signals are then further amplified and read out with a dedicated 20-bit electronic system~\cite{Baulieu:2022} at room temperature.
Data were acquired in continuous mode with a sampling frequency of 100~kHz. Events are triggered offline as described in Sec.~\ref{sec:processing}. 

In RUN014, each detector had an optical fiber running from a cryostat feedthrough at room temperature to a dedicated opening on the top of the detector holder.
The optical fiber was used to illuminate the detector with photons from a laser diode operated at room temperature with a pulse generator and laser driver~\cite{Aereodiode:2019}. 
Two laser wavelengths, 1590~nm and 1650~nm, were used during commissioning to study the detector efficiency and the phonon channel linearity. 
In addition, the laser was used regularly for neutralizing the germanium crystal charge state~\cite{Censier:2004}. 
The latter was performed by setting the electrodes to 0~V and flashing the crystal with photons at a high repetition rate. 
This procedure was repeated three times per week for 1~hour. 
After each neutralization, the bias polarity of the NTL electrodes was inverted to minimize surface charge build-up.

Once thermalized, the detectors were operated in a temperature range between 13.5 and 14.4~mK in RUN013.
In RUN014, the detector temperature was stabilized using a heater with a reference thermometer connected to the detector plate, in order to reduce drifts in the phonon channel gain.
The operating temperatures in RUN014 ranged from 12.0 to 15.6~mK.
The larger operating temperature range in RUN014 was caused by cryogenic instabilities, which were triggered by an additional thermal load due to the presence of HDPE sheets between the copper cans. The HDPE sheets absorb infrared radiation from the warmer copper can and release the heat to the colder can holding them. 
In runs beyond RUN014, the thermal load was improved with two changes: (1)~the HDPE sheets were covered with a reflective foil to limit their radiation absorption, and (2)~the detectors were further decoupled from the 1-K HEMT electronics through the removal of the titanium-alloy structure connecting them (Fig.~\ref{fig:photo}, right).

\subsection{Data collection}
\label{sec:data}
RUN013 and RUN014 lasted 45 and 160~days, respectively. This article focuses on key datasets from RUN014 and reports the baseline resolution improvement observed in RUN013 as a result of enhanced vibration isolation in the setup.
The data on disk represent 90.1\% (87.0\%) of the total elapsed time of RUN013 (RUN014). 
The periods of time without data taking correspond to upgrades of the experimental system external to the cryostat, detector neutralizations, and experiment down time.

The germanium detectors were activated with an AmBe neutron source before deployment at the ILL, so that a significant population of $^{71}$Ge is present in the crystal at the start of the runs.
The K-shell and L-shell lines, at 10.37~keV and 1.3~keV respectively, occur uniformly through the volume of the detector and are visible in the \Ricochet germanium detectors for electron recoil calibration.
Because of the 11.43~day half-life of $^{71}$Ge~\cite{Hampel:1985}, the two lines were strong enough to study in detail the detector response to electron recoils during the first two months of RUN014, up to July.
The detector response to neutrons was measured with an in-situ calibration using a $^{252}$Cf source in September. In total, 135~hours of $^{252}$Cf data were acquired with both detectors and an additional 118~hours were collected with RED167 alone.
The detector performance, in terms of resolution and background, was studied for both reactor-on and reactor-off data.
In total, 577~hours (564~hours) of reactor-on data were recorded on RED167 (RED237), along with 1438~hours (1309~hours) of reactor-off data.
Unless otherwise stated, the datasets described in the latter part of this article are those when the muon veto system and its synchronization with the two germanium detectors were fully operational. This corresponds to 155~hours of reactor-on data and 791~hours of reactor-off data on each detector. 
A few datasets are considered to estimate the detector baseline performance (Table~\ref{tab:resolution}) for specific facility upgrades dedicated to noise improvement and study: (1)~24~hours before and 24~hours after the installation of the damping pads between the cryostat frame and its pillars; (2)~21~hours before and 38~hours after the installation of the damping polyurethane foams between the bellows and the pulse-tube frame; (3)~comparison of the baseline resolution with the pulse tube on and off.
The noise studies are completed with an additional 188+119~hours of data recorded with only one detector at a time (RED167 or RED237, respectively) in the reactor-on period at the start of RUN014, before the inclusion of the muon veto in the acquisition.

\section{Data processing}
\label{sec:processing}

This section outlines the offline processing of ionization, phonon, and muon veto data for RUN013 and RUN014. Note that cryogenic calorimeter data mainly follow the offline processing pipeline described in Ref.~\cite{Colas:2023}, while muon veto data are processed separately. 

The charge amplifiers on both ionization channels of the cryogenic calorimeters have the same gain: one is collecting the holes and the other is collecting the electrons. The signals on the two ionization channels are therefore expected to be opposite to each other. In contrast, the common noise has the same sign on both electrodes. 
Thus, we linearly transform the two traces from the top and the bottom electrodes into the sum and the difference of the two.
The sum is expected to mainly sample the common noise (channel labeled $N$) and the difference contains some residual noise and the summed signals of both channels (channel labeled $S$).
The two ionization channels $S$ and $N$ and the phonon channel $P$ are downsampled from 100~kHz to 50~kHz to improve processing speed. 
It has been verified that the downsampling does not impact the baseline energy resolution.

The phonon and ionization pulses are described by analytical models which capture both the physical detector response and their filtering by read-out electronics. 
Phonon pulse templates are obtained from least-squares fits in the frequency domain of raw pulse traces with energies between 15 and 70~keV$_{\mathrm{ph}}$. 
As the phonon channel response is temperature and detector dependent, phonon templates are created for each detector at 12.0~mK, 14.0~mK, 14.2~mK, 15.0~mK and 15.6~mK, corresponding to the most common temperatures regulated on the detector plate in RUN014. In RUN013, since the temperature was not regulated, a 14.3 mK template was applied to regions of stable detector performance.
A single ionization template per detector is applied to the entire dataset.
To properly reproduce the high frequency poles, as required for improved timing resolution (see Sec.~\ref{sec:coincidences}),  the templates for the ionization channel are derived using high energy events from the $^{41}$Ar peak at 1293.6~keV$_{\mathrm{ee}}$. These events are produced in the decay of neutron-activated $^{41}$Ar~\cite{Bhike:2014} in the air surrounding the cryostat when the reactor is on.

The data processing is performed using an analysis windows of 1~s in order to fully contain both the faster (15~ms) and the slower ($\mathcal{O}$(100)~ms) decay time constants of the phonon pulses. 
The ionization events, with a decay time of $\mathcal{O}$(20)~ms, are also fully contained in the analysis window.
Signal variations that are too slow to be correctly sampled in this interval are attenuated by applying a first-order high-pass prefilter with a cutoff frequency of 2~Hz to the streamed data.

\begin{figure}
    \centering
    \includegraphics[width=1.0\linewidth]{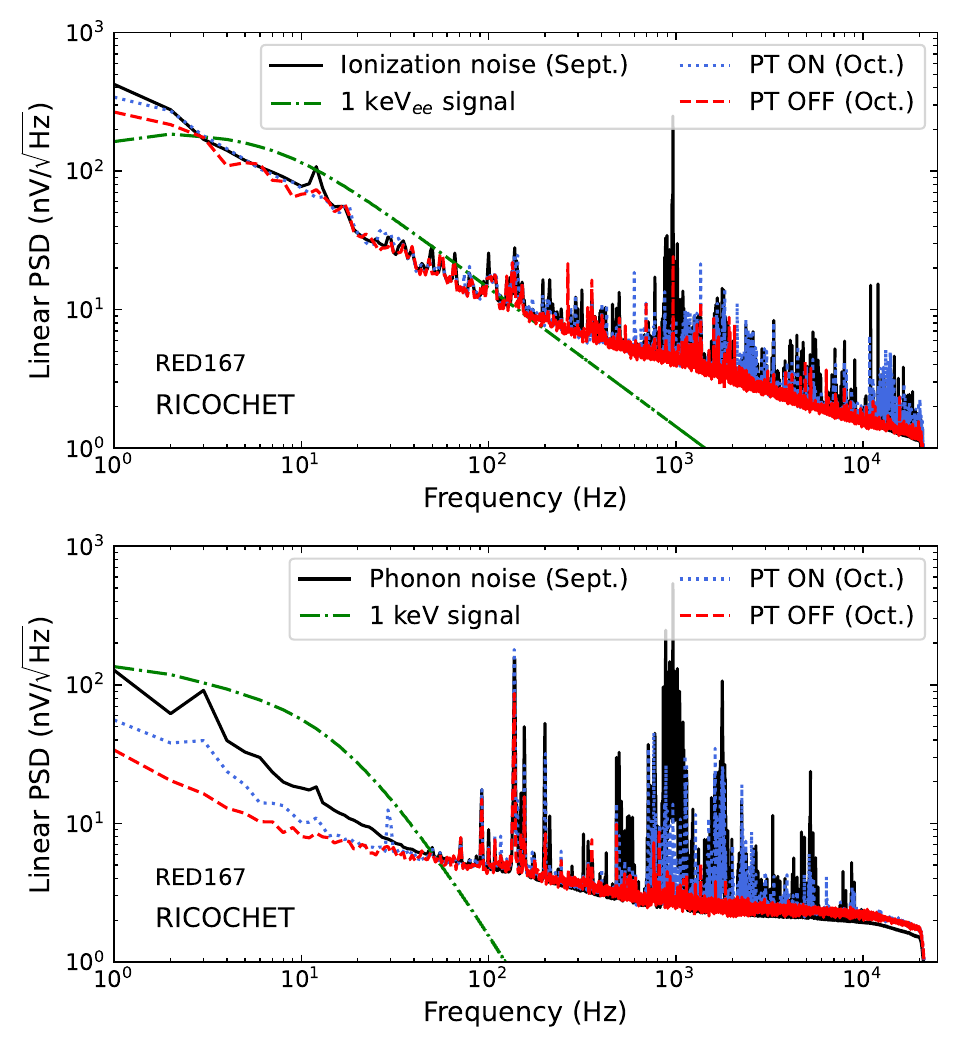}
    \caption{Black curve: noise power spectrum (referred to the input) of the ionization (top) and phonon (bottom) channels for the detector RED167, averaged over the last 118~hours of the $^{252}$Cf calibration (84768 traces recorded in September).
    Blue dotted line (red dashed line): noise power spectra recorded in October, after the addition of the damping polyurethane foam between the bellows and the pulse-tube frame while the pulse tube was in operation (was off). 
    Green dotted-dashed line: norm of the Fourier transform of a 1~keV phonon signal model and a 1~keV$_{\mathrm{ee}}$ ionization signal model.
    See Table~\ref{tab:resolution} for the corresponding baseline resolutions.
    }
    \label{fig:psd}
\end{figure}

Both triggering and pulse amplitude estimation require an accurate noise model. This is determined by evaluating the power spectral density (PSD) of the noise in each channel every 15~minutes, using approximately 180 randomly sampled 1-second traces, which have minimal correlation to the signal model.
In addition, the ionization noise traces are used to evaluate, at each frequency $f$, the average correlation $h_f=\langle N_fS_f \rangle / \langle S_fS_f \rangle$~\cite{Augier:2024} between the noise in the $S$ and $N$ channels.
In the subsequent steps of the analysis, the $S$ channel is replaced by a copy $S'$ where the correlated noise with the $N$ channel is removed by considering $S'_f=S_f-h_fN_f$.
Examples of the noise PSDs for the ionization $S'$ and phonon $P$ channels are shown in Fig.~\ref{fig:psd}, where they are compared with that of a signal with an energy of 1~keV. 

Events are triggered on the basis of the phonon channel information. For this purpose, a matching filter is constructed from the phonon noise and the analytical phonon signal. The trigger algorithm finds the largest filtered phonon channel sample in a 15-minute stream, tags its location, defines a one-second dead-time region around it, and reiterates to find the next largest deviation in the remaining stream, until no more samples are available. This strategy ensures that low-energy events produce a trigger only if the corresponding 1-s trace is devoid of higher-energy pile-up events.

For each triggered event, the precise time and amplitudes of the phonon and ionization signals are obtained by  performing a least-squares fit in the frequency domain of the data in the $P$, $S'$ and $N$ channels. The fit uses the analytical templates and noise PSD described above. The minimized $\chi^2$ is that of the sum of the contribution of the three channels. A common time is imposed between the phonon and ionization signals. The start time of the pulse is scanned in steps of 2~$\upmu$s using a $\pm$3~ms window around the time found by the triggering algorithm. 
This procedure greatly improves the time resolution of events with an ionization signal: despite bandwidth limitations of about 40~kHz from the ionization amplifier, its rise time is still about three orders of magnitude faster than the phonon channel. 
For each triggered event, the fit thus yields one event time (relative to the common clock reference), the ionization and phonon amplitudes, and the contributions of each channel to the total $\chi^2$. 

In order to continuously monitor the contribution of noise to the ionization and phonon baseline resolutions, a least-squares fit is also performed on the empty traces that have been used to evaluate the noise PSD. In these fits, the event time of the pulses is fixed to the center of the window, resulting in a Gaussian distribution of amplitudes. The quoted resolutions in Sec.~\ref{sec:performance} correspond to one standard deviations of these Gaussians.

The muon veto raw data are processed using a separate pipeline that tags muon events and retrieves their time. 
To reduce the $\gamma$ background, which can trigger the muon veto by accidental or correlated coincidence, an energy threshold is applied on both triggered panels. 
The threshold is defined individually for each panel by fitting the muon energy distribution with an empirical model (Landau function convoluted with a Gaussian) and is set at half of the muon peak energy.
This threshold is typically around 3~MeV for the top panels and around 4~MeV for the side ones. 
The muon peak distribution is fit every hour to adjust for gain variations.
The muon detection efficiency with this selection was measured for each panel with a dedicated setup, composed of two small scintillator plates on either side of the main panel and used as a trigger~\cite{Chemin:2024}. The measured efficiency was greater than 98\% for all panels prior to the installation on-site. Using cosmic events, the stability of the efficiency has been monitored on-site to be better than 1\%.

\section{Data analysis}
\label{sec:analysis}
The data analysis is divided in four parts: detector calibration (Sec.~\ref{sec:calibration}), data quality selections (Sec.~\ref{sec:cuts}), trigger and data selection efficiencies for reactor-on and reactor-off data (Sec.~\ref{sec:efficiency}) and coincidences between the germanium detectors and the muon veto (Sec.~\ref{sec:coincidences}).

\subsection{Detector calibration}
\label{sec:calibration}
\begin{figure}
    \centering
    \includegraphics[width=1.0\linewidth]{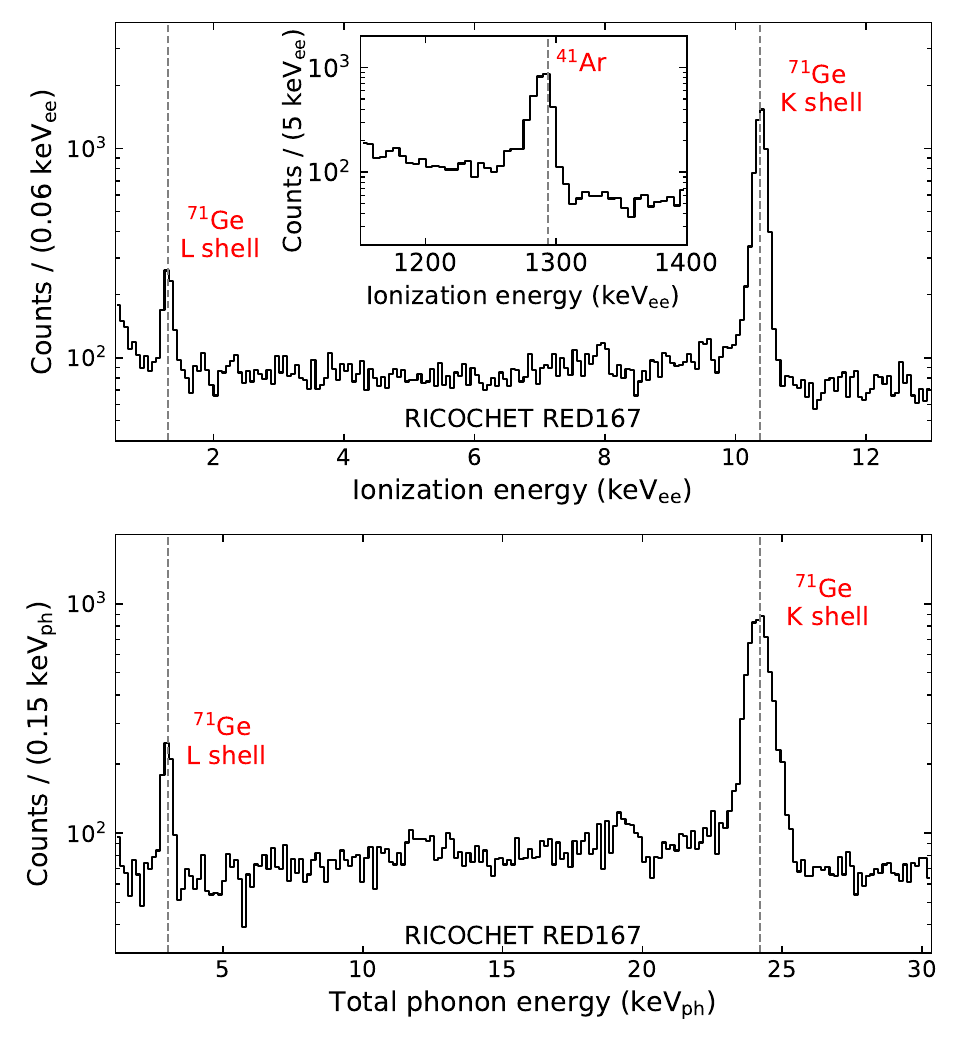}

    \caption{Energy distribution of the ionization (top) and phonon (bottom) channels showing the $^{71}$Ge K-shell and L-shell lines at 10.37~keV and 1.3~keV during the reactor-on period. The inset shows the ionization energy distribution of the $^{41}$Ar line at 1293.6~keV. The vertical dashed lines indicate the expected peak positions assuming a linear calibration for ionization, and after the nonlinearity correction of the phonon channel (see text). 
    The energy distribution of the phonon channel only shows events with at least 0.2~keV$_{\mathrm{ee}}$ ionization signal, removing heat-only events.}
    \label{fig:hist_ion}
\end{figure}

\begin{figure}
    \centering
    \includegraphics[width=1.0\linewidth]{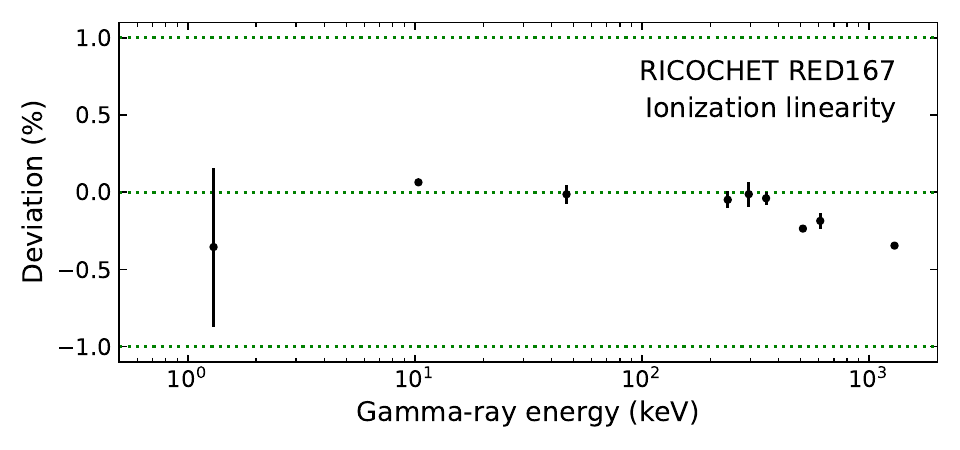}
    \includegraphics[width=1.0\linewidth]{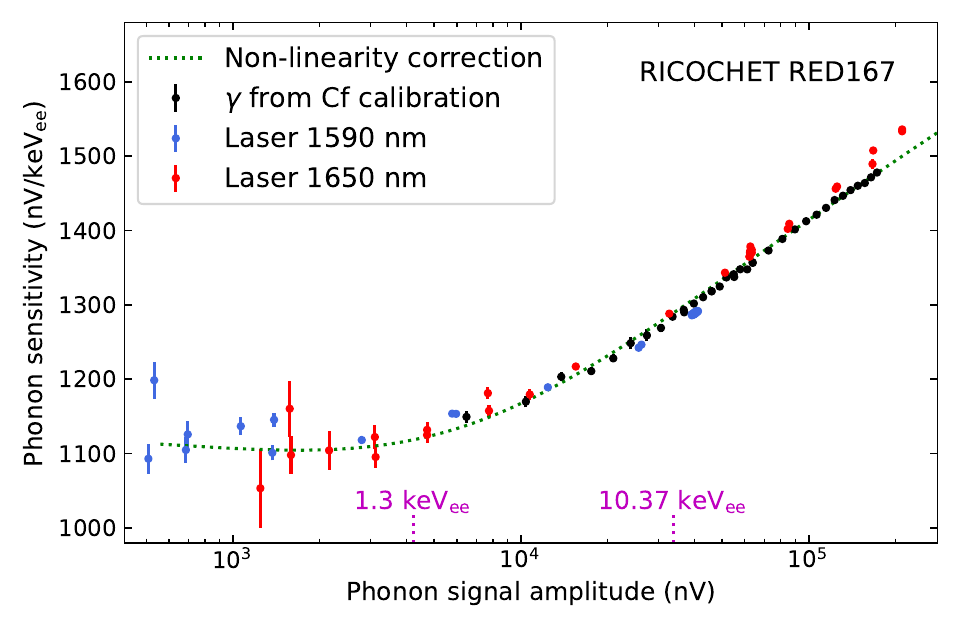}
    \caption{Top panel: deviation from linearity of the ionization signal as a function of energy, as measured from the positions of peaks from $^{71}$Ge, $^{41}$Ar (reactor-on data) and U/Th decay chains (reactor-off data). 
    Bottom panel: Ratio of the amplitude of the phonon pulses (in nV) to the charge signal (calibrated in keV$_{\mathrm{ee}}$) as function of the amplitude of the phonon pulses. The datasets are the electron recoils from the $^{252}$Cf calibration (black), and from the laser pulses at wavelength of 1590~nm (blue) and 1650~nm (red). The uncertainties are statistical only. 
    The curve represents the function used to correct the nonlinearity of the phonon channel, as described in the text. }
    \label{fig:nonlin}
\end{figure}

The ionization and phonon channels are linearly calibrated using, when available, the atomic relaxation cascades produced by the electron capture on the K shell of $^{71}$Ge~\footnote{The contribution of $^{68}$Ge is considered to be negligible for the current runs, given the expected activation rate in Ref.~\cite{Armengaud:2017a} and the detector mass.}, resulting in energy deposits of 10.37~keV.
The maximum daily excursion of the calibrated 10.37~keV line measured during the reactor-on period is used as an indicator of the stability. 
The latter was found to be of 0.3\% (0.4\%) for RED167 (RED237) over 400~hours in the ionization channel.
The linearity of the ionization channel is estimated using the 1.3-keV L-shell line when available, and with the $^{41}$Ar $\gamma$ line at 1293.6~keV. Spectra of these peaks are shown in Fig.~\ref{fig:hist_ion} (top). 
The nonlinearity of the ionization channel for RED167  and RED237 are $-0.4(5)$\%  and $+0.14(16)$\% at 1.3~keV, and $-0.35(2)$\%  and $-0.38(2)$\% at 1293.6~keV. 
The test of nonlinearity is also extended to the 511~keV e$^+$e$^-$ annihilation peak and the most visible $\gamma$ lines from the thorium and uranium chains (Fig.~\ref{fig:nonlin}, top). The nonlinearity on the ionization energy scale is found to be negligible within a 1\% precision requirement and is neglected in the following analyses.

When the K-shell and L-shell lines are not visible above the noise floor, the stability of the ionization channel and its linearity are used to infer the calibration of the phonon channel by imposing that the ratio between the phonon channel and the ionization channel for electron recoil events is equal to $1+q\cdot V/\epsilon$~\cite{Armengaud:2017}, the expected value according to Eq.~\ref{eqn:boost}.

The phonon response has a strong nonlinearity with energy as shown in Fig.~\ref{fig:nonlin} (bottom). The ratio of the phonon amplitude in nV and the energy of the corresponding ionization signal calibrated in keV$_{\mathrm{ee}}$ varies as a function of the amplitude of the pulse. 
This effect is observed for electron recoils (those from the $^{252}$Cf calibration are shown in Fig.~\ref{fig:nonlin}, bottom). 
A very similar effect is also observed for laser events. Dedicated streams were acquired where the laser pulse length was adjusted to produce signals ranging roughly from 100~eV$_\mathrm{ph}$ to 140~keV$_\mathrm{ph}$. 
The ratio starts to deviate from a constant value for signals with energies of the order of 1~keV$_{\mathrm{ph}}$. 
This dependence is parametrized with the empirical function:
\begin{equation}
f(E)=p_0+p_1(\ln{E}-p_2) \dfrac{1+p_4 (\ln{E}-p_2)}{1+\exp \left(\dfrac{\ln{E}-p_2}{p_3} \right)}\;,
\end{equation}
where $E$ is the energy expressed in keV$_\mathrm{ph}$ and $p_{[0,...,4]}$ are free parameters which were determined by a fit to the high statistics sample of $\gamma$ events from the $^{252}$Cf neutron source. The resulting function is used to correct the phonon signal amplitudes and is found to be applicable for most of RUN014.
Slightly different corrections are obtained depending on cryogenic conditions, for example in periods when the germanium detectors are not fully thermalized and the Ge-NTD thermistors have not reached their nominal sensitivities.
In that regime, the difference in temperature between the detectors and the regulated heat bath to which they are connected decreases with time. 
This variation of the detector temperature as a function of time produces a time-dependent gain, which is accounted for by correcting  the anticorrelation between the phonon gain and the pre-pulse mean baseline~\cite{Alessandrello:1998}.
Figure~\ref{fig:hist_ion} (bottom) shows the energy distribution below 30.3~keV$_{\mathrm{ph}}$ for the phonon channel after the linearity and pre-pulse mean baseline corrections.
After selecting the 10.37~keV line in the ionization channel, the maximum daily excursion of the 10.37~keV events measured by the phonon channels during the reactor-on period of RED167 (RED237) is 0.6\% (1.4\%) over 24~days (19~days).

The peaks shown in Fig.~\ref{fig:hist_ion} are characterized by a tail on the left, which is due to partially collected charges.
For both detectors, less than 1\% of the charges in the 10.37~keV peak suffer from significant charge degradation, to the point where events leak down to the nuclear recoil band.

\subsection{Data selections}
\label{sec:cuts}

\begin{figure}
    \centering
    \includegraphics[width=1.0\linewidth]{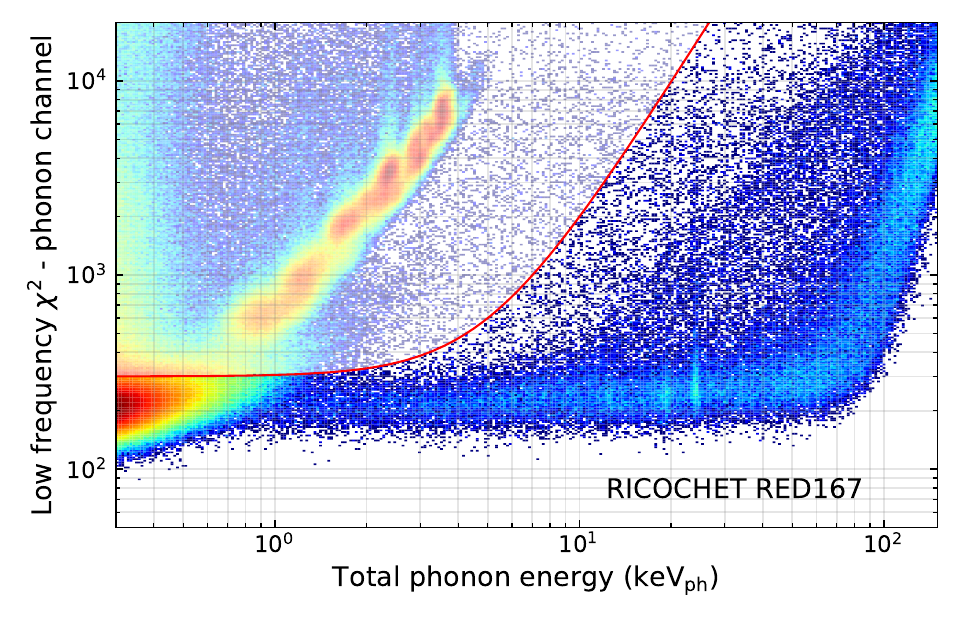}
    \caption{Low-frequency $\chi^2$ of the phonon channel as a function of the total phonon energy for the reactor-on and reactor-off data. The red curve highlights the position of the $\chi^2$ data selection for the phonon channel. Two lines are visible at 18.9 and 24.2~keV$_{\mathrm{ph}}$ corresponding to the copper fluorescence line and the $^{71}$Ge K-shell line after the NTL amplification.}
    \label{fig:chi}
\end{figure}

A series of selection criteria were developed for this dataset.
A total of 15~minutes of data from the reactor-on dataset and 45~minutes from the reactor-off dataset are removed because of reconstruction issues during these periods. 
Two data quality selections are applied to remove events that deviate from the expected behavior for electron and nuclear recoil interactions. 
The data quality of the ionization channel is ensured by removing outliers of the ionization pre-pulse mean baseline. This data selection removes 2.1\% (1.2\%) of data for RED167 (RED237).
The quality of the phonon traces is checked by comparing them to the phonon template below a frequency of 100~Hz.
This quantity is referred to as the low-frequency $\chi^2$. Its distribution as a function of the total phonon energy is shown in Fig.~\ref{fig:chi} for the combined reactor-on and reactor-off data of RED167. The accumulation of events with high $\chi^2$ values and energies below 5~keV$_\mathrm{ph}$ is mostly associated with electromagnetic perturbation of the electronics. The energy dependence of the data selection to remove these events (red line on Fig.~\ref{fig:chi}) is chosen to preserve a high efficiency for good quality 1.3 and 10.37~keV events (i.e. events reconstructed with the correct energy scale within 1\% and having no degradation of the signal/background ratio). 
This data selection also acts as a pile-up rejection. In the reactor-on period, 6.8\% (7.5\%) of the 10.37~keV events for RED167 (RED237) are removed by this selection. 

\subsection{Efficiency}
\label{sec:efficiency}
\begin{figure}
    \centering
    \includegraphics[width=1.0\linewidth]{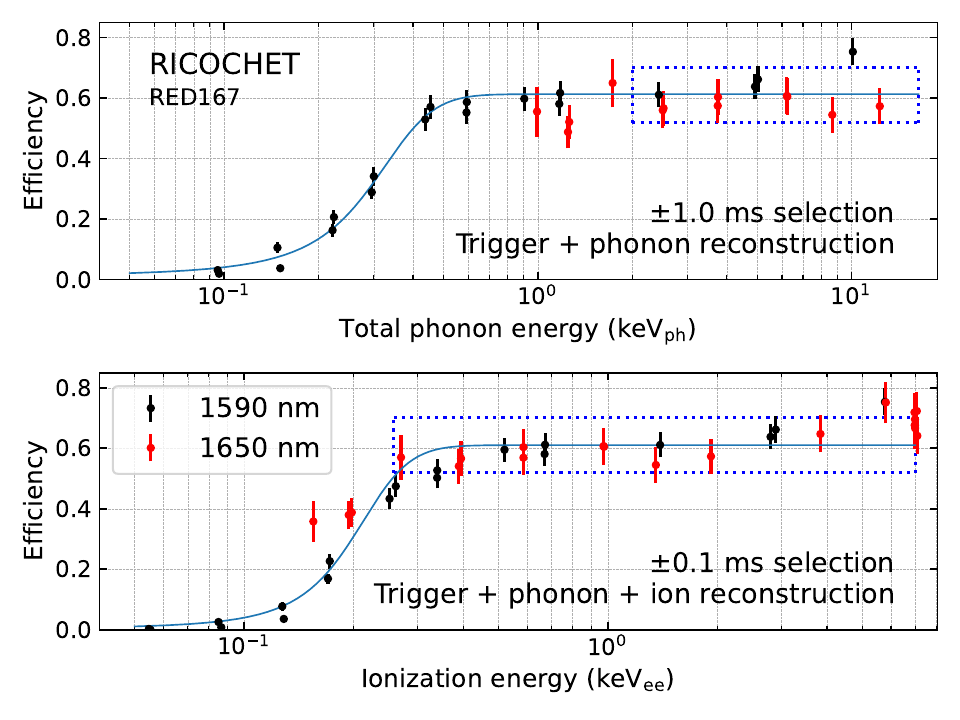}
    \caption{Top panel: Fraction of laser pulse events that are triggered and reconstructed within $\pm1$~ms of the expected pulse time, as a function of the total phonon energy. The points in black (red) are obtained with laser pulses with a wavelength of 1590 (1650)~nm. The uncertainties are statistical only. Bottom panel: the same, but with a stricter $\pm0.1$~ms matching window to ensure a proper reconstruction of the ionization pulse. 
    The data shown in both panels were acquired during a reactor-off period.
    The dotted blue boxes represent the obtained efficiency of 61(9)\%, covering the recoil energy range between 2 and 7~keV for all event categories. 
    This corresponds to the total phonon energy range from 2 to 16.3~keV$_\mathrm{ph}$ and the ionization energy range between 0.25 to 7~keV$_\mathrm{ee}$.
    }
    \label{fig:effic}
\end{figure}

The combined efficiency of the trigger and reconstruction algorithms is measured as the ratio of reconstructed laser events passing the low-frequency $\chi^2$ and ionization pre-pulse mean baseline event selections to the known number of injected laser pulses. 
This measurement as a function of energy performed during the reactor-off period is shown in Fig.~\ref{fig:effic}. The number of expected pulses is given by $fT$, where $f$ is the frequency of the trigger pulses and $T$ is the length in time of the relevant data stream.
The measurement was performed at two different wavelengths (1590 and 1650~nm corresponding to 0.780 and 0.751~eV, respectively) near the germanium band gap, and thus corresponding to two different regimes of penetration of the infrared light in the crystal~\cite{Lattaud:2024}.
A phonon pulse is considered as a valid match if it was within $\pm$1~ms from the expected laser pulse.
The phonon energy of events with such delays is precisely reconstructed. 
This is not the case for the much faster ionization signal, which requires a stricter matching window of $\pm$0.1~ms.
Invalid ionization reconstruction of valid phonon pulses occurs only for ionization energies below 0.3~keV$_{\mathrm{ee}}$.
These measurements were performed at a repetition rate of 0.1~Hz to avoid biasing the noise estimation.
The restricted number of injected pulses prevents the study of systematic effects below the size of the statistical uncertainties (about 10\%) at each energy and is not as precise as the pulse injection procedure via software described in Refs.~\cite{Colas:2023_thesis, Armatol:pipeline}. 
However, it provides an estimate of the absolute efficiency of 61(9)\% in the region of interest for the RED167 data in Fig.~\ref{fig:effic}, and 63(9)\% for the equivalent data for RED237. 
The relative systematic uncertainty of 15\% on these values is sufficiently precise for the evaluation of the different background rates observed in the present commissioning stage, in the recoil energy range from 2 to 7~keV (covered by the blue dashed box on Fig.~\ref{fig:effic}).
These plateau values are obtained from the fit of a sigmoid function to all data in Fig.~\ref{fig:effic}. This same fit also indicates that half of the plateau is reached at an ionization energy of 0.2~keV$_{\mathrm{ee}}$ for both detectors, ensuring that the efficiency is compatible with the energy range covered by the boxes on Fig.~\ref{fig:effic} (bottom). The required energy range for the phonon energy is well within the flat plateau region.
 
The stability in time of the efficiency throughout the run was tested by continuously injecting $\sim$15~keV$_{ee}$ laser pulses at a rate of 1/60~Hz. 
The efficiency in the reactor-on period was observed to be systematically reduced by 15\% due to increased pile-up events, mostly due to the $^{41}$Ar background. 
The pulse injection results in an additional dead time of 1.7\% in all periods.
The efficiency of RED167 for the reactor-on and reactor-off periods are thus taken as 52(8)\% and 60(9)\%, respectively. The corresponding values for RED237 are 54(8)\% and 62(9)\%, respectively.

\begin{figure}
    \centering
    \includegraphics[width=1.0\linewidth,trim={0.5cm 0 1.3cm 1.2cm},clip]{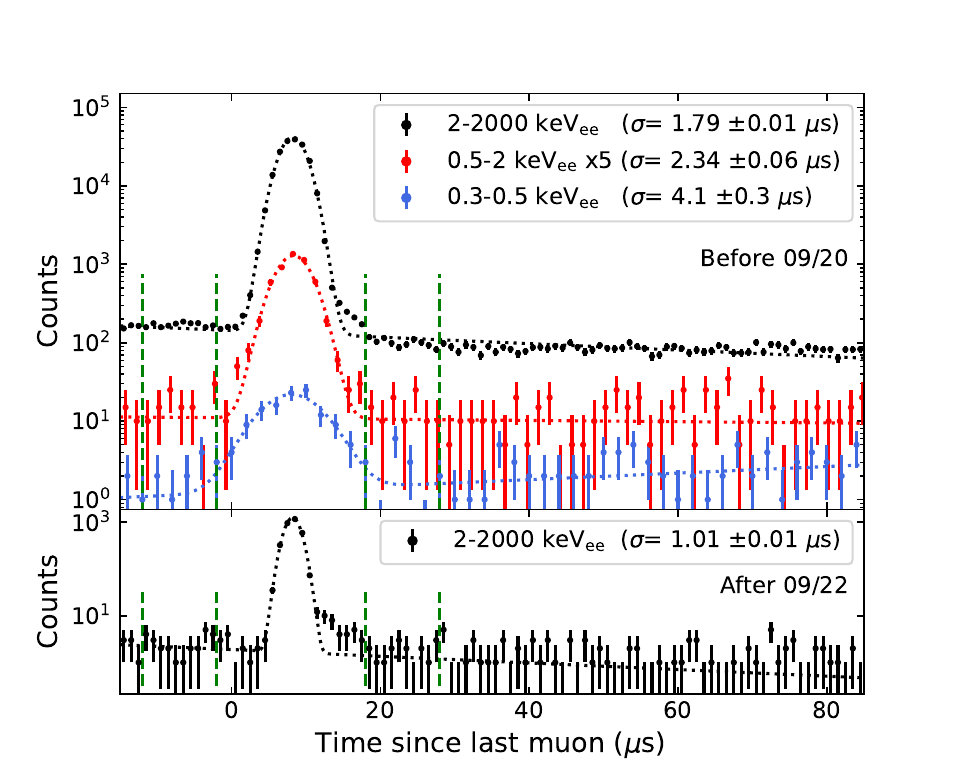}
    \caption{Histogram of the time difference between hits in the muon veto and in the germanium detectors for four event categories.  Top: events taken before September 20$^{\mathrm{th}}$ for different ionization energy ranges: 0.3--0.5~keV$_{\mathrm{ee}}$ (blue),  0.5--2~keV$_{\mathrm{ee}}$ (red, rate multiplied by 5 for clarity) and 2--2000~keV$_{\mathrm{ee}}$ (black). Bottom panel: all events between 2 and 2000~keV$_{\mathrm{ee}}$ taken after the clock firmware update on September 21$^{\mathrm{st}}$. The $\sigma$ values are obtained from the fits shown as dotted lines. The dashed green lines represent the muon selection and side bands discussed in the text.
     }
    \label{fig:muonveto}
\end{figure}

\subsection{Muon veto coincidence}
\label{sec:coincidences}
The cryogenic calorimeters were operated together with the outer muon veto system for part of RUN014. 
When the muon veto is operating, a time difference with the closest muon event can be computed for each event based on its ionization signal which exhibits a much better timing resolution than the phonon channel.
Figure~\ref{fig:muonveto} shows the distribution of the time difference used to identify coincident events: events with a time difference smaller than $\pm$10~$\upmu$s around the peak are tagged as coincident with the muon veto.
The black points on the two panels in Fig.~\ref{fig:muonveto} compare the time difference distribution for events with an ionization energy between 2~keV$_{\mathrm{ee}}$ and 2.0~MeV$_{\mathrm{ee}}$ acquired before (top) and after (bottom) an upgrade of the firmware of the synchronization card, which provided an improvement of the timing resolution from 1.79(1)~$\upmu$s to 1.01(1)~$\upmu$s. This upgrade does not affect the data reported in this article.
The blue and red data in Fig.~\ref{fig:muonveto} (top) show the time difference distribution for all events with an ionization energy from 0.3 to 0.5~keV$_{\mathrm{ee}}$ and from 0.5 to 2~keV$_{\mathrm{ee}}$, respectively, which demonstrates that the majority of coincident events below 2~keV$_{\mathrm{ee}}$ remain within the $\pm10~\upmu$s selection used in Sec.~\ref{sec:bkg}.

\section{Detector resolution performance}
\label{sec:performance}
\begin{table*}
\caption{\label{tab:resolution}%
Baseline energy resolution ($\sigma$) of the ionization and phonon channels measured when the detectors were operated in different conditions: reactor on/off, presence of neutron source, pulse tube on/off, and one or two detector operation. For each data type, the status of the reactor and pulse tube are listed, together with the total number of hours considered (Duration), the month when the data were acquired and the number of detectors operated at the same time (Det.). 
All the data of the same type are fit together with a Gaussian. The uncertainty reported in parentheses is the quadratic sum of the statistical and systematic uncertainties, the latter being obtained by comparing the resolution obtained from a fit of the entire dataset to the average of the values obtained when the data are segmented sequentially in groups of 10$^4$~noise traces. The curly brackets indicate the observed rms spread, when significant, that can be attributed to time variations among the subsamples.
    }
    \begin{ruledtabular}
        \begin{tabular}{lccccccccc}
            Data type&Reactor&Pulse& Month&Duration& Det.&
            \multicolumn{2}{c}{Ionization (eV$_\mathrm{ee}$)}&\multicolumn{2}{c}{Phonon (eV$_\mathrm{ph}$)}\\
            ~&~&Tube&of 2024&(hour)&&RED167&RED237&RED167&RED237\\
            \colrule
Reactor on & on & on & May & 119 &1&- & 43.2(4)\{1\} & -& 95(2)\{1\} \\
Reactor on & on & on & Jun. & 188 &1& 45.7(4) & -& 114(2)\{10\} &-   \\
Reactor on        & on  & on  & Jul.     & 155 &2& 54.3(6)\{3\}  & 46.5(5)\{4\} & 80(1)\{4\} & 106(2)\{1\}\\
Reactor off       & off & on  & Jul./Aug.   & 791 &2& 44.0(4)\{7\} & 43.3(4)\{30\} & 79(1)\{8\} & 108(2)\{8\}\\
$^{252}$Cf source & off & on & Sep. & 135 &2& 42.2(4) & 40.1(1)\{3\} & 72(1)\{6\}& 79(1)\{2\} \\
$^{252}$Cf source & off & on  &   Sep.    &118 &1& 40.2(3) & - & 66(1)\{4\} & -\\
Pulse tube on     & off & on  & Oct.   &0.5 &1& 38(2) & - & 57(3) & -\\
Pulse tube off    & off & off & Oct.   &0.25&1& 44(3) & - & 50(4) & -\\
        \end{tabular}
    \end{ruledtabular}
\end{table*}

The \cenns region of interest for germanium is between the threshold and below 2~keV and therefore a good parameter to quantify the detector performance is the baseline energy resolution.
The baseline resolution is measured by the 1$\sigma$ dispersion of the amplitudes fit to the noise traces. 
First, all baseline noise data are fit together to extract the reference values reported in Table~\ref{tab:resolution}, then the data are segmented sequentially in groups of 10$^4$~noise traces and fit individually to estimate their variation in time (curly brackets in Table~\ref{tab:resolution}).
The difference between the reference value and the average of the set of group resolutions is added to the statical uncertainty and reported in parentheses in Table~\ref{tab:resolution}.
Table~\ref{tab:resolution} summarizes the baseline resolutions obtained in different configurations during RUN014.
Eight cases are compared, corresponding to different conditions: reactor on or off, with or without the presence of the $^{252}$Cf source (in each case, with the operation of either a single detector or two detectors simultaneously), and with or without the pulse tube in operation. 
Except for the two last cases, each measurement spans an interval of more than 100~hours. These measurements are spread over six months during which the mechanical decoupling of the cryostat from the environment was gradually improved. 
It should be observed that the germanium detectors require approximately one month to reach full thermal equilibrium with the bath. The phonon channel resolution improves over this period.

The ionization and phonon resolutions are found to depend on three important factors: (1)~the simultaneous operation of one or two detectors, (2)~the increased disturbances (i.e. vibrations, acoustic noise level) when the reactor is on, for example, from the operation of pumps, and (3)~the quality of the decoupling of the detector from vibrations.

During the reactor-on period, an increase in ionization resolution between 3 and 8~eV$_{\mathrm{ee}}$ is observed when two detectors are readout simultaneously. This is caused by crosstalk noise from neighboring readout modules. The electromagnetic shielding has been improved in the final version of the readout electronics which does not show any signs of significant crosstalk anymore.

The impact of vibrations and their mitigation measures had already been observed in RUN013, where it was found that the installation of damping pads between the cryostat frame and its pillars improved the baseline resolution for the heat (ionization) channel by 35(1)\% (5.3(5)\%) while the reactor was on. 
In RUN014, the installation of the damping polyurethane foam between the bellows and the pulse-tube frame at the end of September resulted in an improvement in the baseline resolution by 20(5)\% for the phonon channel and with no change in the ionization channel. 
The effect of this added insulation on the vibrations caused by the reactor operation has not yet been tested, but Table~\ref{tab:resolution} shows that those caused by the pulse tube are significantly reduced.
After the foam installation, the baseline resolutions measured with the pulse tube on and off are compatible within 2$\sigma$ (last two lines in Table~\ref{tab:resolution}).

The overall improvement in resolutions with time in Table~\ref{tab:resolution} is an encouraging sign that the setup upgrades between July and October should reduce the effect of vibrations in future reactor-on runs.

\section{Background}
\label{sec:bkg}

\begin{figure}
\centering
\includegraphics[width=1.0\linewidth]{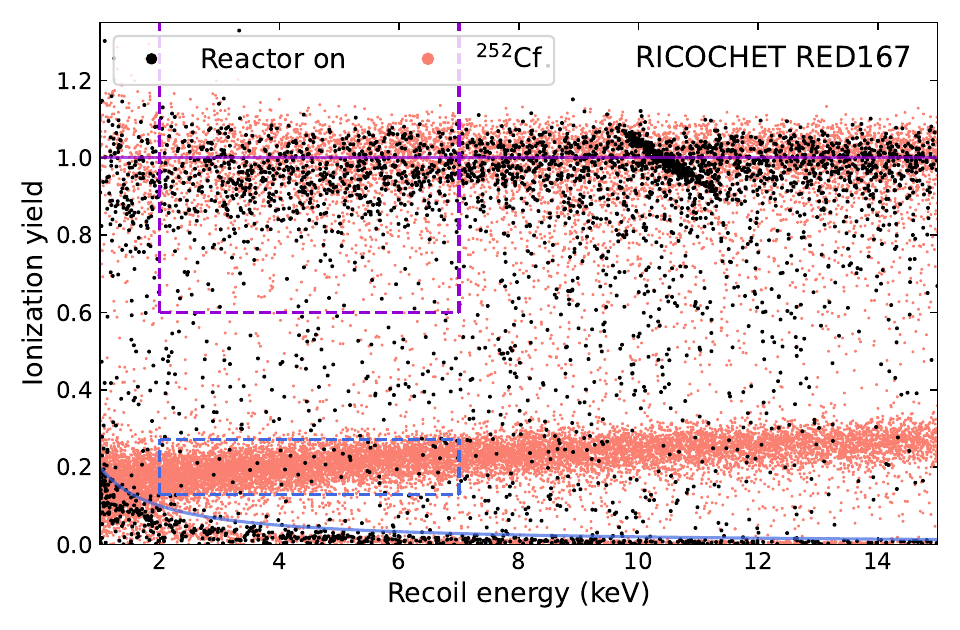}
\caption{Ionization yield as a function of the recoil energy for the reactor-on data in anticoincidence with the muon veto (black) superimposed to the $^{252}$Cf neutron source data (pink). The neutron source data highlight the expected position of the nuclear recoil band. The expected position of electron recoil events and the 0.2~keV$_\mathrm{ee}$ ionization selection used to identify heat-only events are indicated with a violet and blue solid lines, respectively. The violet and blue dashed boxes indicate the ranges covered for the electron recoil and nuclear recoil rates in Table~\ref{tab:bkg}, respectively.}
\label{fig:loglog}
\end{figure}

\begin{table*}[t]
\caption{\label{tab:bkg} Average event rates after correction for efficiency in events per kg, per day and per keV (DRU) in the energy range from 2 to 7~keV for: the heat-only (HO) band, the electron recoil (ER) band, all counts in the nuclear recoil (NR) band, and the NR estimate from the likelihood fit model described in the text. Uncertainties include the statistical uncertainty and the systematic uncertainties on the efficiency. Upper limits have a 90\% C.L., deduced from the likelihood fit described in the text. The backgrounds are reported both for the reactor on and off in anticoincidence with the muon veto (periods in July and August in Table~\ref{tab:resolution}) and are compared to the rates extracted from the \Ricochet GEANT4 simulations of cosmogenic (cosm. sim.) and reactogenic (react. sim.) backgrounds in the same energy range. In addition the reactor-off data are reported with muon veto coincidence selection to quantify the nuclear recoil background induced by muons. An exposure of 155~hours (791~hours) is used for the reactor-on (reactor-off) data.
}
\begin{ruledtabular}
    \begin{tabular}{llcccccccc}
        \textrm{Reactor}&
        \textrm{$\upmu$ coinc./}&
        \textrm{Detector}&
        \textrm{HO (DRU)}&
        \multicolumn{2}{c}{\textrm{ER (DRU)}}&
        \textrm{All NR band (DRU)}&
        \multicolumn{3}{c}{\textrm{NR estimate (DRU)}}\\
        \textrm{ }&
        \textrm{anticoinc.}&
        \textrm{ }&
        \textrm{data}&
        \textrm{data}&\textrm{cosm. sim.}&
        \textrm{data}&
        \textrm{data}&\textrm{cosm. sim.} &\textrm{react. sim.}\\
        \colrule
    \multirow{2}{3em}{on} 
        & \multirow{2}{4.2em}{anticoinc.}
        & RED167 & 707(110) & 1115(172) & 23(2) & 70(10) & $<$ 25 & 4.6(8) & 0.4(1)\\
        &
        & RED237 & 784(122) & 1200(184) & 23(2) & 54(9) & $<$ 9 & 4.6(8) & 0.4(1)\\
    \multirow{2}{3em}{off}
        & \multirow{2}{4.2em}{anticoinc.}
        & RED167 &  642(97) &135(21) & 23(2) & 69(4) & $<$ 15 & 4.6(8) &  -\\
        &
        & RED237 & 611(92) & 123(19) &23(2) &  72(4) & $<$ 21 & 4.6(8) & -\\
    \multirow{2}{3em}{off}  
        & \multirow{2}{4.2em}{coinc.}
        & RED167 & - & 328(50) & 308(7) &  23(2) & 14(3) & 15(2) & -\\
        &
        & RED237 & - & 304(46) & 308(7) & 24(2) & 13(3) & 15(2) & -\\
    \end{tabular}
\end{ruledtabular}
\end{table*}

Figure~\ref{fig:loglog} shows a scatter plot of the ionization yield $\mathcal{Q}$ as a function of the recoil energy $E_r$ for the July reactor-on data and the $^{252}$Cf neutron calibration data. 
Events within $\pm$10~$\upmu$s of a muon veto coincidence are rejected. 
This window was chosen to cover the prompt coincidences of muon veto and germanium detector signals of $>0.3$~keV$_\mathrm{ee}$ (muon selection band in Fig.~\ref{fig:muonveto}) and to minimize random coincidences within the coincidence window.
Applying the anticoincidence selection results in a dead time of $\sim$1\%, which is taken into account in Table~\ref{tab:bkg}.
The energy resolutions of the germanium detectors for these data samples are listed in Table~\ref{tab:resolution}.
The \cenns signal is expected to appear below 2~keV in the nuclear recoil (NR) band. This band is delineated by the dense pink population at $\mathcal{Q}\sim 0.2$ from $^{252}$Cf data on Fig.~\ref{fig:loglog}.
The other two most visible bands in this figure correspond to electron recoil events (ER), centered at $\mathcal{Q}=1$, and a population of events with no significant ionization signal, corresponding to the heat-only (HO) population.
In the present commissioning phase, the objective is to assess the current background levels in the recoil energy range from 2 to 7~keV, where the interpretation of the ionization yield is the most straightforward and relatively insensitive to the variation of detector performance observed as the experimental setup was tested and improved.
In addition, this choice excludes \cenns events from the reactor-on data, events from the $^{71}$Ge L-shell line at 1.3~keV and the copper fluorescence line at 8.1~keV.
The high-statistics $^{252}$Cf data sample makes possible a determination in this energy range of the mean and standard deviation of the measured ionization yield values for NR events, $\mathcal{Q}_n$ and $\sigma_{\mathcal{Q}_n}$.
Figure~\ref{fig:loglog} also shows that in this range, these three populations can be separated by simple selections on the ionization yield for ER and NR events (dashed boxes corresponding to $\mathcal{Q}>0.6$ and $\mathcal{Q}_n\pm 2\sigma_{\mathcal{Q}_n}$, respectively), or by requiring $E_{ion}<0.2$~keV$_{\mathrm {ee}}$ for HO events (blue solid lines), i.e. approximately 5$\sigma_{ion}$, according to Table~\ref{tab:resolution}.

This figure shows a fourth population lying outside the narrow bands where ER, NR and HO events are expected, with ionization yield values rather evenly distributed between 0 and 1.
These events are likely electron recoils with incomplete charge collection. 
This population will be referred as ``surface'' events since this behavior is expected for events near the surface, such as $\beta$ decays on the surfaces of the detector or its copper holder. 
Another source are the electrons produced in photoelectric and Compton interactions in the copper holder. 
The observation of such a high rate in the present planar detector data prevents a clear identification of NR events over the apparent background of surface events in the reactor-on and reactor-off datasets.
The full \Ricochet payload will include detectors with two electrode designs: planar and Fully Inter-Digitized (FID), described in Ref.~\cite{Armengaud:2017}. 
The FID detectors will be capable of rejecting surface background events because of their electrode design.
The observation of surface events has prompted the study of a protocol for etching~\cite{Hoppe:2007} the detector holder surfaces for planar detectors, in addition to the existing cleaning procedure. 

The ER and HO rate in the recoil energy range from 2 to 7~keV can be reliably derived from the observed number of events with ionization yield values above 0.6 and ionization signal below 0.2~keV$_{\mathrm {ee}}$, respectively. 
The reactor-on and reactor-off rates for the two detectors in anticoincidence with the muon veto and the reactor-off rates for muon-coincident events are also listed in Table~\ref{tab:bkg}.
The rates are corrected for the efficiency obtained in Sec.~\ref{sec:analysis}, i.e. 60(9)\% and 62(9)\% in reactor-off periods, and 52(8)\% and 54(8)\% in reactor-on periods for RED167 and RED237, respectively.
The side bands shown in Fig.~\ref{fig:muonveto} are used to correct the event rates for random coincidences.
Muon-related rates were not corrected for barometric pressure variations or changes in water level in the different sections of the transfer channel as these corrections would be smaller than the uncertainties on the efficiencies.
The rates coincident with a muon veto event are consistent between reactor on and reactor off. Only the reactor-off rates are reported in Table~\ref{tab:bkg} due to the larger exposure of the dataset.

The HO rates are similar in both detectors, regardless of whether the reactor is on or off. They did not vary in time during the entire period when the veto was fully operational. 
Incidents where this rate suddenly increased -- notably after hardware intervention on the cryostat -- are under investigation.
However, the rate of HO events returns to nominal after a thermal cycling and no such incidents happened to coincide with the periods covered by Table~\ref{tab:bkg}.
Despite the unknown source of these events, the relative stability of their rates suggests that they can be effectively taken into account via the frequent reactor-on/reactor-off modulation provided by the ILL. 

The ER anticoincident rates of the two detectors are very similar, but increase by more than a factor of 7 when the reactor is turned on.
This increase is likely due in large part to the significant rate of $\gamma$ rays from the decay of $^{41}$Ar~\cite{Abele:2025} ($T_{1/2}$ = 109.61(4)~minutes), as attested by the appearance of a strong characteristic peak at 1293.6~keV (Fig.~\ref{fig:hist_ion}, inset) when the reactor is on.
The large increase in ER rates is not accompanied by a higher rate in the NR band. This indicates that the leakage of low-energy Compton events down to the NR band is compatible with the $<1$\% limit obtained for events in the 10.37~keV peak (see Sec.~\ref{sec:analysis}).
Both the ER and HO populations appear to be clearly separated from the NR band, at least within the uncertainties imposed by the presence of surface events.

\begin{figure*}
    \centering
    \includegraphics[width=1\linewidth]{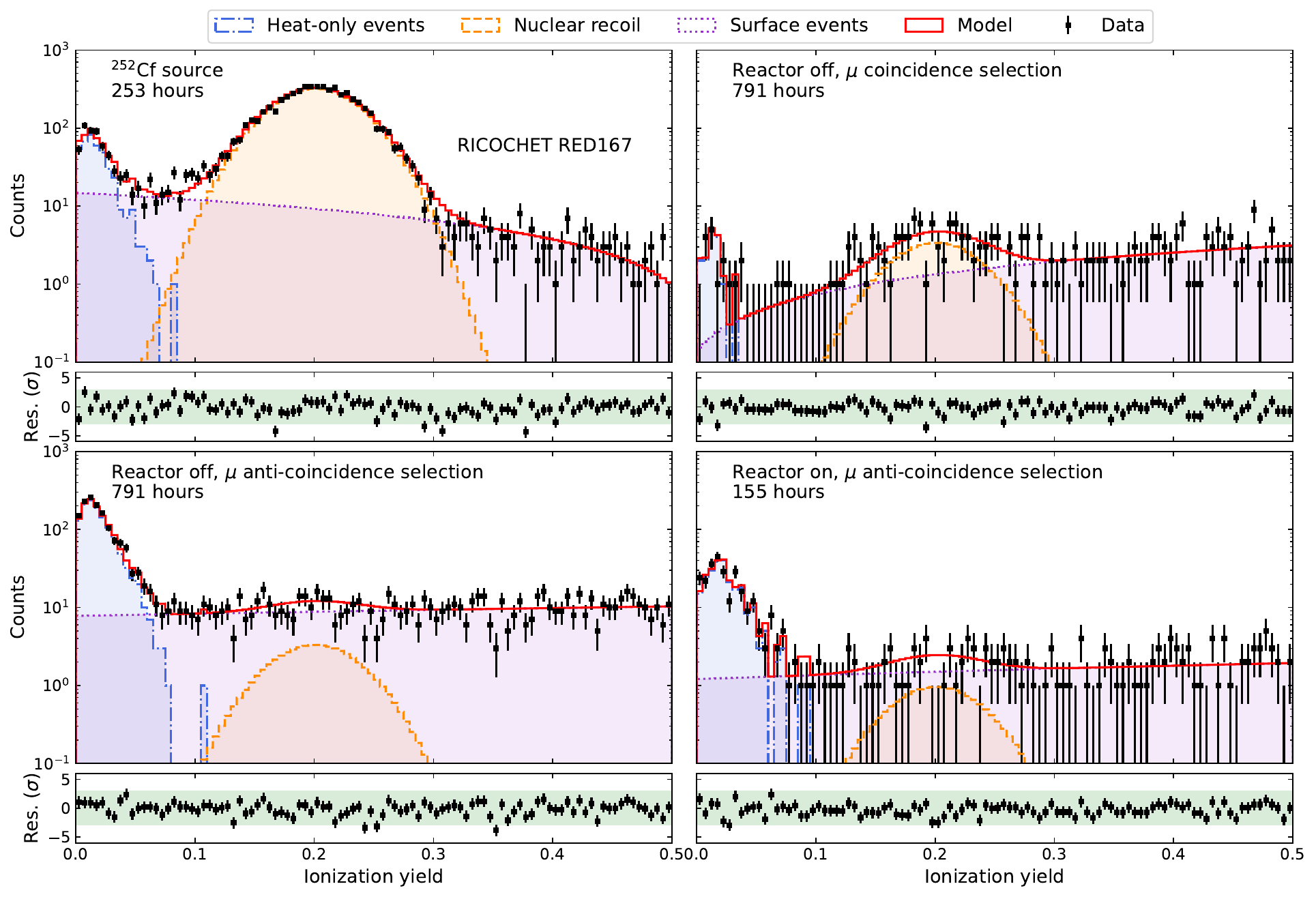}
    \caption{Distribution of the ionization yield in the region between 2 and 7~keV. The data (black) are fit with the model (red), which is composed of three components: heat-only events (blue), nuclear recoils (orange), which are modeled with the $^{252}$Cf source data, and surface events (violet). The residual between the data and the model is reported below each plot with a $\pm$3$\sigma$ band (green) highlighted.
    Top panel: $^{252}$Cf source data (left) and reactor-off data occurring in coincidence with muon events (right). 
    Bottom panel: reactor-on data in anticoincidence with the muon veto (left) and reactor-off data in anticoincidence with the muon veto (right). }
    \label{fig:Qhist}
\end{figure*}

Because of the presence of surface events, the number of background counts in the NR band only provides an overestimation of the NR rate.
To distinguish between the contribution of surface and neutron events in the nuclear recoil band in the range from 2 to 7~keV, a likelihood analysis was performed with all events with ionization yield values between 0 and 0.5.
The resulting fits are reported in Fig.~\ref{fig:Qhist} for four types of data: $^{252}$Cf neutron calibration data (top-left), reactor-off data in coincidence (top-right) and anticoincidence (bottom-left) with the muon veto, and reactor-on data in anticoincidence with the muon veto (bottom-right).
The respective live times of these samples (corresponding to calibrated data being available for both cryogenic calorimeters and muon detectors) are listed in Table~\ref{tab:resolution}.
Nuclear recoil events are modeled using a Gaussian with a mean and sigma set by the $^{252}$Cf data. 
The surface background is modeled as a slope in $\mathcal{Q}$, as ($a\cdot \mathcal{Q} + b$), where $a$ and $b$ are free parameters. 
In the Poisson likelihood function, the rates of nuclear recoil and surface background events are free parameters. The HO event distribution is determined from the sample of events with ionization energies below 0~keV$_\mathrm{ee}$, assuming that the distribution of this variable for HO events is symmetric around zero.
The efficiencies during the reactor-on and reactor-off periods described in Sec.~\ref{sec:efficiency} are used as nuisance parameters.

The binned likelihood fits of samples in Fig.~\ref{fig:Qhist} (top) provide NR rates which are significantly above zero. The rates obtained in Fig.~\ref{fig:Qhist} (bottom) are consistent with zero and a 90\% C.L. limit on the NR rate was extracted instead.
The rate measurements and upper limits obtained in the likelihood analysis are listed in the ``NR estimate'' column of Table~\ref{tab:bkg}. 
The ``All NR band'' column reports the total rates obtained in the $\pm 2 \sigma_{\mathcal{Q}_n}$ NR band. 
For the muon anticoincidences, these rates are a factor three larger than the NR limits in the ``NR estimate'' column. 
This emphasizes the importance of reducing the population of surface events, either by the use of FID detectors or with more rigorous surface treatments. 

The operation of the ILL reactor and of the experiments in the surrounding hall induces a significant ambient neutron flux.
However, no significant NR excess is observed in the reactor-on data, within the $\sim$10~events/(kg\,day\,keV) sensitivity of the present measurements and limits.
A combined 90\% C.L. upper limit of 9~NR~events/(kg\,day\,keV) was obtained for the reactor-on data between 2 and 7~keV.
This constitutes the first test in real conditions of the lead and HDPE shields, the muon veto, and of the particle identification capability of the mini-CryoCube detectors.

The NR rate in coincidence with the muon veto allows us to measure the NR background induced by cosmic rays (modulo the geometric and detection efficiency of the veto). 
These events will be rejected in the \cenns analysis, but their measurement provides an important check of the \Ricochet cosmogenic neutron GEANT4 simulations described in Ref.~\cite{Augier:2023a} that have been used to validate the design of the lead and HDPE shields, as well as the muon veto geometry.
The cosmogenic background was generated using the CRY generator~\cite{Hagmann:2007, CRY} with a complete description of the ILL building and applying a flux normalization tuned with outdoor measurements performed on-site~\cite{Allemandou:2018}. 
An equivalent of 10~days of such simulated cosmic-ray events is used in this article. Refined with respect to Ref.~\cite{Augier:2023a}, the response of the muon veto is simulated according to the performance obtained in the qualification campaign of the muon detectors performed prior to installation, while the true deposited energy is used as the response of the germanium detectors. The muon trigger conditions in terms of thresholds and configuration are also reproduced in the simulation. 

The simulation estimates ER and NR rates in coincidence with the muon trigger of 308(7) and 15(2)~events/(kg\,day\,keV) respectively, in the range from 2 to 7~keV.
They are in excellent agreement with the measurements within their uncertainties as it is visible in Table~\ref{tab:bkg}.
The calculated rates of NR events in anticoincidence with the muon veto, 4.6(8)~events/(kg\,day\,keV) from cosmogenic and 0.4(1)~events/(kg\,day\,keV) from reactogenic simulations, are below the measured limits.
These simulations do not include radiogenic background sources nor the presence of activated $^{41}$Ar, therefore they are not expected to reproduce the ER rate measured in anticoincidence.

\section{Conclusion}
\label{sec:conclusions}
The \Ricochet experiment aims to measure the \cenns for reactor antineutrinos at the Institut Laue-Langevin.
The experimental setup was installed at the end of 2023 and commissioned in 2024. The detector performance and background levels of two 42-gram germanium cryogenic calorimeters with a heat and ionization readout are reported in reactor-on and reactor-off conditions.
The resolution performance improved as different vibration mitigation measures were incorporated to the experimental setup. 
At the end of commissioning, an ionization baseline resolution of 40~eV$_\mathrm{ee}$ was achieved for both detectors and 50~eV$_\mathrm{ph}$ (80~eV$_\mathrm{ph}$) with 4~V applied across the detector electrodes on the RED167 (RED237) phonon channel.

The experimental background was measured in anticoincidence with the muon veto system during both the reactor-on and reactor-off periods.
Different sources of backgrounds were resolved using the particle identification capabilities of the \Ricochet detectors.
A combined upper limit of 9~nuclear recoil events/(kg\,day\,keV) at 90\% C.L. was obtained during the reactor-on period between 2 and 7~keV.
A population of events consistent with a near-surface interaction was identified as the most important background contribution in the nuclear recoil band. 
Two approaches will be undertaken for its mitigation: (1)~the use of detectors with a Fully Inter-Digitized electrode geometry capable of identifying surface events, which is currently being tested at the ILL, and (2)~the etching of the copper holders for planar detectors is being investigated to reduce their surface contamination level.
An additional reduction in the cosmogenic background is expected from the operation of a cryogenic muon veto that will increase the solid-angle coverage. 
Improvements of the shielding airtightness are being considered in order to reduce the $^{41}$Ar background and reduce the dead time in reactor-on periods.

The electron recoil and nuclear recoil backgrounds coincident with muon veto events were also measured.
A good agreement between these backgrounds and the \Ricochet GEANT4 simulation was found. 
While these backgrounds will be rejected with the muon veto, its measurement allowed us to validate the GEANT4 simulations of the cosmogenic background.

The results presented in this article from the commissioning runs suggest that \Ricochet is well-positioned to be the first cryogenic-calorimeter-based \cenns experiment located at a nuclear reactor to enter its science phase.

\begin{acknowledgments}
Authors are grateful for the technical and administrative support of the ILL for the installation and operation of the \Ricochet detector.
We acknowledge CC-IN2P3 for providing and maintaining computing and storage services needed for the experiment.
We are grateful to O.~Bourrion (LPSC) and D.~Tourres (LPSC) for the synchronization of the muon veto acquisition. 
This project received funding from the European Research Council (ERC) under the European Union’s Horizon2020 research and innovation program under the Grant Agreement ERC-StGCENNS 803079, the French National Research Agency (ANR) within the project ANR-20-CE31-0006, the project ANR-22-EXES-0001, the LabEx Lyon Institute of Origins (ANR-10-LABX-0066) of the Universit\'e de Lyon, within the Plan France2030, and the NSF under Grants PHY-2209585, PHY-2013203 and PHY-2411390, Natural Sciences and Engineering Research Council of Canada (NSERC), the Canada First Excellence Research Fund, and the Arthur B. McDonald Institute (Canada). The HEMT production and development was supported in part by the French network RENATECH. A portion of the work carried out at MIT was supported by DOE QuantISED award DE-C0020181 and the Heising-Simons Foundation. This work is also partly supported within the State Project ``Science'' by the Ministry of Science and Higher Education of the Russian Federation (075-15-2024-541).
\end{acknowledgments}

\bibliographystyle{apsrev4-1} 
\bibliography{bibliography}
\end{document}

%% file: authors.tex
\providecommand{\affiliation}[1]{#1}
\newcommand{\JINR}{\affiliation{Department of Nuclear Spectroscopy and Radiochemistry, Laboratory of Nuclear Problems, JINR, Dubna, Moscow Region, Russia 141980}}
\newcommand{\ILUSA}{\affiliation{Department of Physics and Astronomy, Northwestern University, IL, USA}}
\newcommand{\MAUSA}{\affiliation{Department of Physics, University of Massachusetts at Amherst, Amherst, MA, USA 01003}}
\newcommand{\ON}{\affiliation{Department of Physics, University of Toronto, ON, Canada M5S 1A7}}
\newcommand{\CSGRENOBLE}{\affiliation{Institut Laue-Langevin, CS 20156, 38042 Grenoble Cedex 9, France}}
\newcommand{\MAUSATwo}{\affiliation{Laboratory for Nuclear Science, Massachusetts Institute of Technology, Cambridge, MA, USA 02139}}
\newcommand{\COUSA}{\affiliation{Physics Department, Colorado School of Mines, Golden, CO, USA 80401}}
\newcommand{\CNRSINPF}{\affiliation{Univ Lyon, Universit\'e Lyon 1, CNRS/IN2P3, IP2I-Lyon, F-69622, Villeurbanne, France}}
\newcommand{\CNRSINP}{\affiliation{Univ. Grenoble Alpes, CNRS, Grenoble INP, Institut N\'eel, 38000 Grenoble, France}}
\newcommand{\CNRSINPLPSCINP}{\affiliation{Univ. Grenoble Alpes, CNRS, Grenoble INP, LPSC-IN2P3, 38000 Grenoble, France}}
\newcommand{\CNRSCN}{\affiliation{Univ. Paris-Saclay, CNRS, C2N, Palaiseau, 91120 Palaiseau, France}}
\newcommand{\CNRSINPTwo}{\affiliation{Universit\'e Paris-Saclay, CNRS/IN2P3, IJCLab, 91405 Orsay, France}}

\author{A.~Armatol}\CNRSINPF
\author{C.~Augier}\CNRSINPF
\author{L.~Bailly-Salins}\CNRSINPLPSCINP
\author{G.~Baulieu}\CNRSINPF
\author{L.~Berg\'e}\CNRSINPTwo
\author{J.~Billard}\CNRSINPF
\author{J.~Bl\'e}\CNRSINPLPSCINP
\author{G.~Bres}\CNRSINP
\author{J-.L.~Bret}\CNRSINP
\author{A.~Broniatowski}\CNRSINPTwo
\author{M.~Calvo}\CNRSINP
\author{A.~Cavanna}\CNRSCN
\author{A.~Cazes}\CNRSINPF
\author{E.~Celi}\ILUSA
\author{D.~Chaize}\CNRSINPF
\author{M.~Chala}\CNRSINPLPSCINP
\author{M.~Chapellier}\CNRSINPTwo
\author{L.~Chaplinsky}\MAUSA
\author{G.~Chemin}\CNRSINPLPSCINP
\author{R.~Chen}\ILUSA
\author{J.~Colas}\CNRSINPF
\author{L.~Couraud}\CNRSCN
\author{E.~Cudmore}\email[Corresponding author: ]{elspeth.cudmore@mail.utoronto.ca}\ON
\author{M.~De~Jesus}\CNRSINPF
\author{N.~Dombrowski}\MAUSATwo
\author{L.~Dumoulin}\CNRSINPTwo
\author{A.~Durnez}\CNRSCN
\author{O.~Exshaw}\CNRSINP
\author{S.~Ferriol}\CNRSINPF
\author{E.~Figueroa-Feliciano}\ILUSA
\author{J.~A.~Formaggio}\MAUSATwo
\author{S.~Fuard}\CSGRENOBLE
\author{J.~Gascon}\CNRSINPF
\author{A.~Giuliani}\CNRSINPTwo
\author{C.~Goy}\CNRSINPLPSCINP
\author{C.~Guerin}\CNRSINPF
\author{E.~Guy}\CNRSINPF
\author{L.~Haegel}\CNRSINPF
\author{S.~A.~Hertel}\MAUSA
\author{C.~Hoarau}\CNRSINPLPSCINP
\author{Z.~Hong}\ON
\author{J.-C.~Ianigro}\CNRSINPF
\author{Y.~Jin}\CNRSCN
\author{A.~Juillard}\CNRSINPF
\author{T.~Khussainov}\thanks{also at: Institute of Nuclear Physics of the Ministry of Energy of the Republic of Kazakhstan, 1 Ibragimov Street, 050032, Almaty, Kazakhstan}\JINR
\author{A.~Kubik}\thanks{also at: SNOLAB, Creighton Mine \#9, 1039 Regional Road 24, Sudbury, ON P3Y 1N2, Canada}\ON
\author{J.~Lamblin}\CNRSINPLPSCINP
\author{H.~Lattaud}\CNRSINPF
\author{T.~Le-Bellec}\CNRSINPF
\author{L.~Leroy}\CNRSCN
\author{M.~Li}\MAUSATwo
\author{A.~Lubashevskiy}\thanks{also at: Lebedev Physical Institute of the Russian Academy of Sciences, 53 Leninskiy Prospect, 119991, Moscow, Russia}\JINR
\author{S.~Marnieros}\CNRSINPTwo
\author{N.~Martini}\CNRSINPF
\author{J.~Minet}\CNRSINP
\author{A.~Monfardini}\CNRSINP
\author{F.~Mounier}\CNRSINPF
\author{V.~Novati}\email[Corresponding author: ]{valentina.novati@lpsc.in2p3.fr}\CNRSINPLPSCINP
\author{E.~Olivieri}\CNRSINPTwo
\author{P.~K.~Patel}\MAUSA
\author{E.~Perbet}\CNRSINPLPSCINP
\author{H.~D.~Pinckney}\MAUSATwo
\author{D.~V.~Poda}\CNRSINPTwo
\author{D.~Ponomarev}\thanks{also at: Lebedev Physical Institute of the Russian Academy of Sciences, 53 Leninskiy Prospect, 119991, Moscow, Russia}\JINR
\author{W.~Van~De~Pontseele}\COUSA
\author{J.-S.~Real}\CNRSINPLPSCINP
\author{F.~C.~Reyes}\email[Corresponding author: ]{freyes@mit.edu}\MAUSATwo
\author{A.~Rodriguez}\ILUSA
\author{M.~Rousseau}\CNRSINPLPSCINP
\author{S.~Rozov}\JINR
\author{I.~Rozova}\JINR
\author{B.~Ryan}\MAUSATwo
\author{D.~Sabhari}\ILUSA
\author{S.~Scorza}\CNRSINPLPSCINP
\author{R.~Serra}\CSGRENOBLE\CNRSINPLPSCINP
\author{Ye.~Shevchik}\JINR
\author{T.~Soldner}\CSGRENOBLE
\author{A.~Stutz}\CNRSINPLPSCINP
\author{Ch.~Ulysse}\CNRSCN
\author{L.~Vagneron}\CNRSINPF
\author{S.~Vasilyev}\JINR
\author{F.~Vezzu}\CNRSINPLPSCINP
\author{P.~Vittaz}\CNRSINPF
\author{E.~Yakushev}\JINR
\author{J.~Yang}\MAUSATwo
\author{D.~Zinatulina}\JINR

\collaboration{\textsc{Ricochet} Collaboration}\noaffiliation